\documentclass[prd,%
 reprint,
 amsmath,amssymb,
 aps,
]{revtex4-2}

\usepackage{graphicx}
\usepackage{dcolumn}
\usepackage{bm}
\usepackage{hyperref}

\footnotetext[1]{Corresponding author: creissel@mit.edu}
\footnotetext[2]{These authors contributed equally to this work}
\footnotetext[3]{Also at Center for Gravitational Physics, University of Texas at Austin, Austin, TX 78712, United States}

\begin{document}

\preprint{APS/123-QED}

\title{Coherence DeepClean: \\ Toward autonomous denoising of gravitational-wave detector data}

\author{Christina Reissel$^{*,\dagger}$}
\affiliation{%
MIT Department of Physics, Cambridge, MA 02139, United States
}%
\affiliation{%
MIT LIGO Laboratory, Cambridge, MA 02139, United States
}%
\affiliation{%
NSF Institute on \textit{Accelerated AI Algorithms for Data-Driven Discovery} (A3D3), \\ MIT, Cambridge, MA 02139 and University of Minnesota, Minneapolis, MN 55455, United States
}%

\author{Siddharth Soni$^{\dagger}$}%
\affiliation{%
MIT Department of Physics, Cambridge, MA 02139, United States
}%
\affiliation{%
MIT LIGO Laboratory, Cambridge, MA 02139, United States
}%


\author{Muhammed Saleem$^{\ddagger}$}
\author{Michael Coughlin}%
\affiliation{%
School of Physics and Astronomy, University of Minnesota, Minneapolis, MN 55455, United States
}%
\affiliation{%
NSF Institute on \textit{Accelerated AI Algorithms for Data-Driven Discovery} (A3D3), \\ MIT, Cambridge, MA 02139 and University of Minnesota, Minneapolis, MN 55455, United States
}%

\author{Philip Harris}
\affiliation{
MIT Department of Physics, Cambridge, MA 02139, United States
}%
\affiliation{%
NSF Institute on \textit{Accelerated AI Algorithms for Data-Driven Discovery} (A3D3), \\ MIT, Cambridge, MA 02139 and University of Minnesota, Minneapolis, MN 55455, United States
}%

\author{Erik Katsavounidis}
\affiliation{
MIT Department of Physics, Cambridge, MA 02139, United States
}%
\affiliation{%
MIT LIGO Laboratory, Cambridge, MA 02139, United States
}%
\affiliation{%
NSF Institute on \textit{Accelerated AI Algorithms for Data-Driven Discovery} (A3D3), \\ MIT, Cambridge, MA 02139 and University of Minnesota, Minneapolis, MN 55455, United States
}%


\date{\today}

\begin{abstract}
Technical and environmental noise in ground-based laser interferometers designed for gravitational-wave observations like Advanced LIGO, Advanced Virgo and KAGRA, can manifest as narrow ($<1$\,Hz) or broadband ($10'$s or even $100'$s of Hz) spectral lines and features in the instruments' strain amplitude spectral density.
When the sources of this noise cannot be identified or removed,
in cases where there are witness sensors sensitive to this noise source, denoising of the gravitational-wave strain channel can be performed in software, enabling recovery of instrument sensitivity over the affected frequency bands.
This noise hunting and removal process can be particularly challenging due to the wealth of auxiliary channels monitoring the interferometry and the environment and the non-linear couplings that may be present.
In this work, we present a comprehensive analysis approach and corresponding cyberinfrastructure to promptly identify and remove noise in software using machine learning techniques.
The approach builds on earlier work (referred to as {\tt{DeepClean}}) in using machine learning methods for linear and non-linear regression of noise.
We demonstrate how this procedure can be operated and optimized in a tandem fashion close to online data taking; it starts off with a coherence monitoring analysis that first singles out and prioritizes witness channels that can then be used by {\tt{DeepClean}}.
The resulting denoised strain by {\tt{DeepClean}} reflects a 1.4\% improvement in the binary neutron star range, which can translate into a 4.3\% increase in the sensitive astrophysical volume.
This cyberinfrastructure we refer to as {\tt{Coherence DeepClean}}, or {\tt CDC}, is a significant step toward autonomous operations of noise subtraction for ground-based interferometers.
\end{abstract}

\maketitle


\section{Introduction}
The field of gravitational-wave physics entered a new era in 2015 when the two Laser Interferometer Gravitational-wave Observatory (LIGO) detectors~\cite{LIGOScientific:2014pky} in the US made the first direct detection of a merging pair of black holes~\cite{LIGOScientific:2016aoc}. 
Since then, the Virgo detector~\cite{TheVirgo:2014hva} in Italy and the KAGRA detector~\cite{KAGRA:2020tym} in Japan have joined the effort of observing the gravitational-wave sky.  
In the first three observing runs from 2015 to 2020 (designated as O1, O2 and O3), the LIGO-Virgo-KAGRA network published a catalog of 90 compact binary coalescence events~\cite{KAGRA:2021vkt}.
This includes signals from binary black holes (BBH), neutron star-black holes (NSBH), and binary neutron stars (BNS). After the O3 run, several improvements were made to the LIGO detectors, including new test mass mirrors and frequency-dependent squeezing~\cite{LIGOO4Detector:2023wmz, Capote:2024rmo}. These upgrades have improved LIGO's sensitivity in the fourth observing run (O4) of the LIGO-Virgo-KAGRA network of detectors~\cite{Capote:2024rmo} that started in May 2023 and is ongoing, with a planned end date in October 2025.

Many noise sources reduce the sensitivity of the instruments on both short and long time scales.  Reducing the fundamental noise sources, such as quantum or thermal noise, requires improved technologies like frequency-dependent squeezing and better materials for test mass optics~\cite{aLIGO:2020wna, LIGOO4Detector:2023wmz, Capote:2024rmo, LIGO:2024kkz, LIGO:2021ppb}. Other than fundamental noise, instrument sensitivity is adversely impacted by short-duration noise transients and long-duration noise couplings originating from the environment or the interferometry~\cite{LIGOScientific:2019hgc, LIGO:2024kkz, LIGO:2021ppb}. For a detailed discussion on transient noise and its impact on the sensitivity of the detectors, we refer the reader to the literature~\cite{LIGO:2024kkz, LIGO:2021ppb}. In this paper, we focus on analyzing and reducing the impact of long-duration (persistent) environmental and instrumental noise couplings that manifest themselves as narrowband or broadband lines and features in the instruments' strain amplitude spectral density. 

Over the last few years, machine learning has started to play a very pivotal role in gravitational-wave data analyses.
This includes astrophysical searches~\cite{Aframe,Gunny:2021gne,Skliris:2020qax,Koloniari:2024kww}, parameter estimation~\cite{Dax:2021tsq,Chatterjee:2024pbj, Williams:2021qyt, Wong:2023lgb}, 
monitoring squeezing~\cite{Whittle:2023lky}, classifying transient noise~\cite{Zevin:2016qwy, Soni:2021cjy, Mukund:2016thr}, removing noise couplings~\cite{PhysRevResearch.2.033066,Saleem:2023hcm, PhysRevD.101.042003, Mukund:2020lby, Wang2024, Chatterjee:2024obg} and multiple other areas~\cite{Cuoco:2020ogp}. The role of machine learning techniques in gravitational-wave data analyses will only increase in the near future. This article presents our work on reducing a variety of long-duration noise in gravitational-wave data, focusing on the Advanced LIGO detectors, one in Livingston, Louisiana, and another one in Hanford, Washington State, with a fully automated pipeline.
Our pipeline combines measurements of the actual instrumental and environmental noise conditions with machine learning denoising algorithms that ultimately improve the instruments' sensitivity to astrophysical sources. 

The paper is organized as follows. We first discuss the types of noise and couplings present in the detectors in section~\ref{sec:Noise}. Next, we describe in detail the process of measuring the coherence between different detectors' channels in section~\ref{sec:Coherence_Monitor}. Following that, we review the {\tt DeepClean} algorithm \cite{PhysRevResearch.2.033066, Saleem:2023hcm} used for noise cleaning and discuss our algorithmic updates to it in section~\ref{sec:Denoising}. Finally, we present our results in section~\ref{sec:Results} and summarize the work in section~\ref{sec:Summary}.

\section{Noise in ground-based gravitational-wave detectors}
\label{sec:Noise}

Long-duration noise in ground-based gravitational-wave detectors is characterized by a complex, composite spectrum consisting of both broadband noise and spectral (narrow) lines. While known contributions to the noise budget at frequencies above $1$\,kHz are dominated by quantum noise originating from fluctuations in the phase quadrature of the vacuum field (shot-noise), dominant contributions at lower frequencies include fundamental noise sources such as coating thermal noise and quantum noise, technical noise sources such as control noise, and environmental sources such as scattered light noise and beam jitter~\cite{Capote:2024rmo}. 
The noise originates from various sources such as the AC power grid, mechanical resonances of the mirror suspensions, and detector control systems. While fundamental noise is irreducible for a given detector technology/setup, technical noise can be reduced via hardware improvements and in software. 

For frequencies below $200~\mathrm{Hz}$, the differences between expected (fundamental) and measured (actual) noise in the present interferometric detectors are not fully understood~\cite{Capote:2024rmo}. Since differences are most apparent in this frequency band, our work focuses on identifying and subtracting noise below 200\,Hz.
Extending this work over the entire spectrum of interest for gravitational-wave detection is straightforward and as we describe in section~\ref{sec:Denoising} this is a mere matter of configuration parameters of our analysis workflow. Subtracting noise leads to increased sensitivity allowing to observe a higher rate and signal-to-noise ratio for gravitational-wave events \cite{Davis:2018yrz}.

Seismometers, microphones, accelerometers, radio receivers, magnetometers, and interferometer state controls are widely used in gravitational-wave detectors to monitor detector behavior and environmental conditions. These auxiliary channels provide important information about possible noise contributions. In total, the detectors monitoring output sums up to thousands of auxiliary or witness channels~\cite{LIGOScientific:2016gtq}.
Previous efforts addressing noise subtraction in strain data using information from auxiliary channels include Wiener filter approaches~\cite{Davis:2018yrz} as well as machine learning extensions to include non-linearities and non-stationarities~\cite{PhysRevD.101.042003,PhysRevD.105.102005}. 
During the fourth observation run (O4), both the LIGO Hanford and LIGO Livingston observatories apply low-latency narrow-line subtraction and cleaning using the NonStationary Estimation and Noise Subtraction (NonSENS) algorithm for dedicated noise couplings leading to an increase in the BNS range between 1-5\,Mpc~\cite{Capote:2024rmo}.

In our analysis, we consider only ``safe" channels. An auxiliary channel is designated as ``safe" if the main gravitational-wave sensing channel does not couple to it ~\cite{Smith:2011an,PhysRevD.103.042003,LIGO:2024kkz,LIGO:2021ppb}
and can therefore be assumed free of potential gravitational-wave signals. The safety of channels is established through routine signal injection tests performed via modulation of the laser power of the photon calibrator~\cite{Karki:2016pht}. The end test masses are displaced in a way that mimics the passage of a gravitational wave through the detector~\cite{LIGO:2024kkz}, which is why we refer to the setup and measurement as hardware injections.
This analysis established around 900 channels as ``safe"~\cite{PhysRevD.103.042003,LIGO:2024kkz,LIGO:2021ppb}
at each detector during the current fourth observing run (O4). Our pipeline focuses on determining the coherence between these ``safe" channels and the strain-recording one. We then use the information to apply a machine learning-based denoising, {\tt DeepClean}.

\section{Coherence Studies}\label{sec:Coherence_Monitor}
Signal coherence is a statistical measure that allows the study of the relationship between two signals. It can be interpreted as the frequency domain analog of the cross-correlation function, a measure of similarity between the two signals. Mathematically, coherence between two signals $x(t)$ and $y(t)$ is given by:

\begin{align}
    \gamma(f) = \frac{|S_{xy}(f)|^2}{S_{xx}(f)S_{yy}(f)}
\end{align}

\noindent
where $S_{xx}$, $S_{yy}$ and $S_{xy}$ are the auto spectral density of $x(t)$, auto spectral density of signal $y(t)$ and cross spectral density of $x(t)$ and $y(t)$ \cite{GARDNER1992113}. 

\begin{figure}
    \centering
    \includegraphics[width=1.0\linewidth]{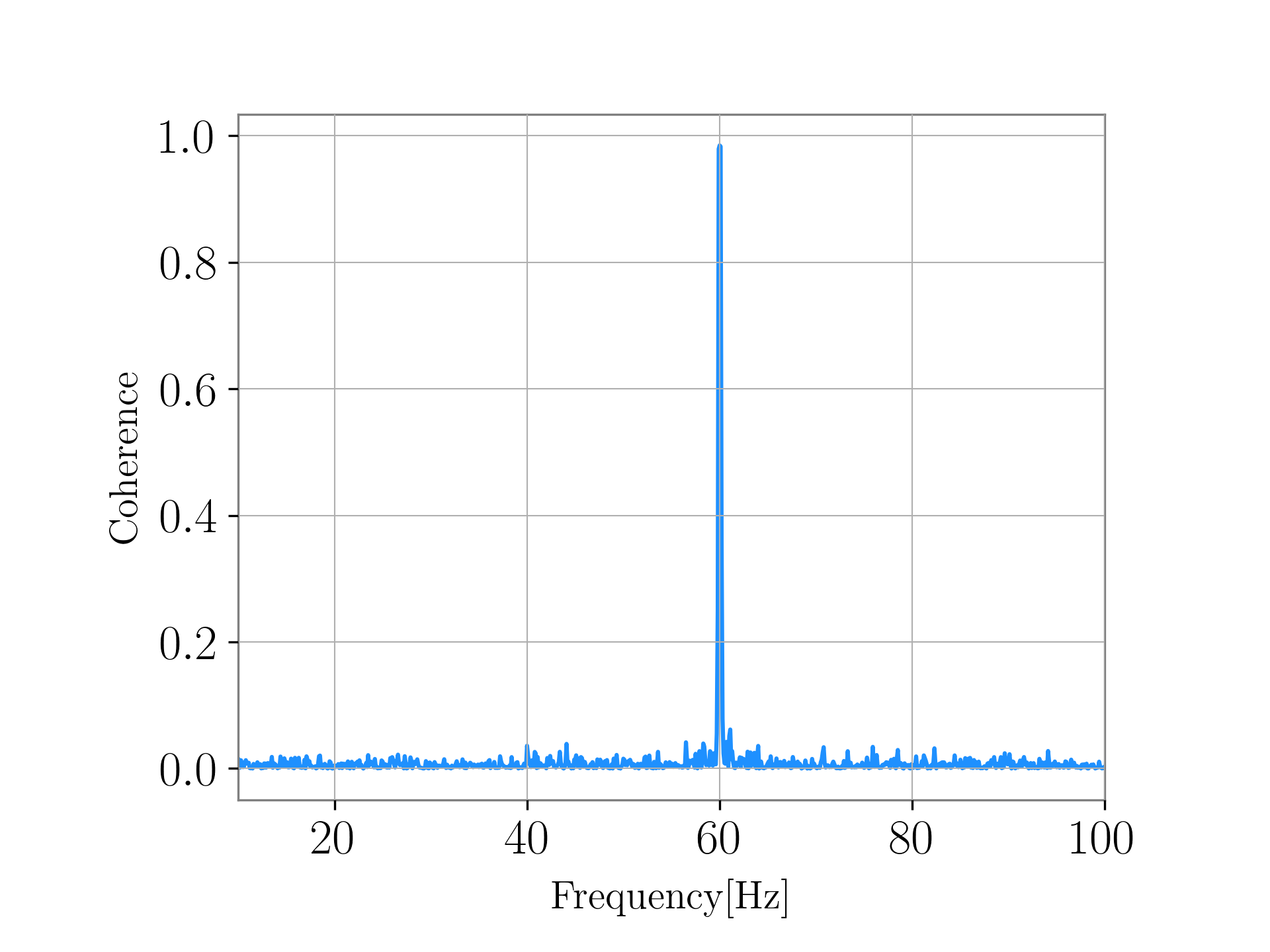}
    \caption{Coherence between h(t) and a LIGO Livingston detector corner station magnetometer. The peak at $60$ Hz is due to the power mains line.}
    \label{fig:mag_coh}
\end{figure}

In a linear system, coherence calculation informs how much of the measured output signal can be explained by the input signal.
It assumes values between 0 and 1 for a given frequency.
At any given frequency, a 0 value indicates no correlation between the input and output while a value of 1 indicates the output signal power can be all accounted for by the input one (signals are fully correlated).
Coherence analysis between channels (time series) that monitor the interferometry and the environment and the strain stream $h(t)$ is an effective way in studying noise couplings to the gravitational-wave measurement. This assumes the channel couplings are roughly linear, such that a coherence metric can determine the correlation between channels. 
Figure \ref{fig:mag_coh} shows the coherence between a LIGO Livingston magnetometer and the strain data between $10$--$100~\mathrm{Hz}$. The peak at $60$ Hz is due to the power line noise coupling \cite{LIGOScientific:2019hgc}. 

\subsection{Coherence monitoring}
\begin{figure*}
    \centering
    \includegraphics[width=0.9\linewidth]{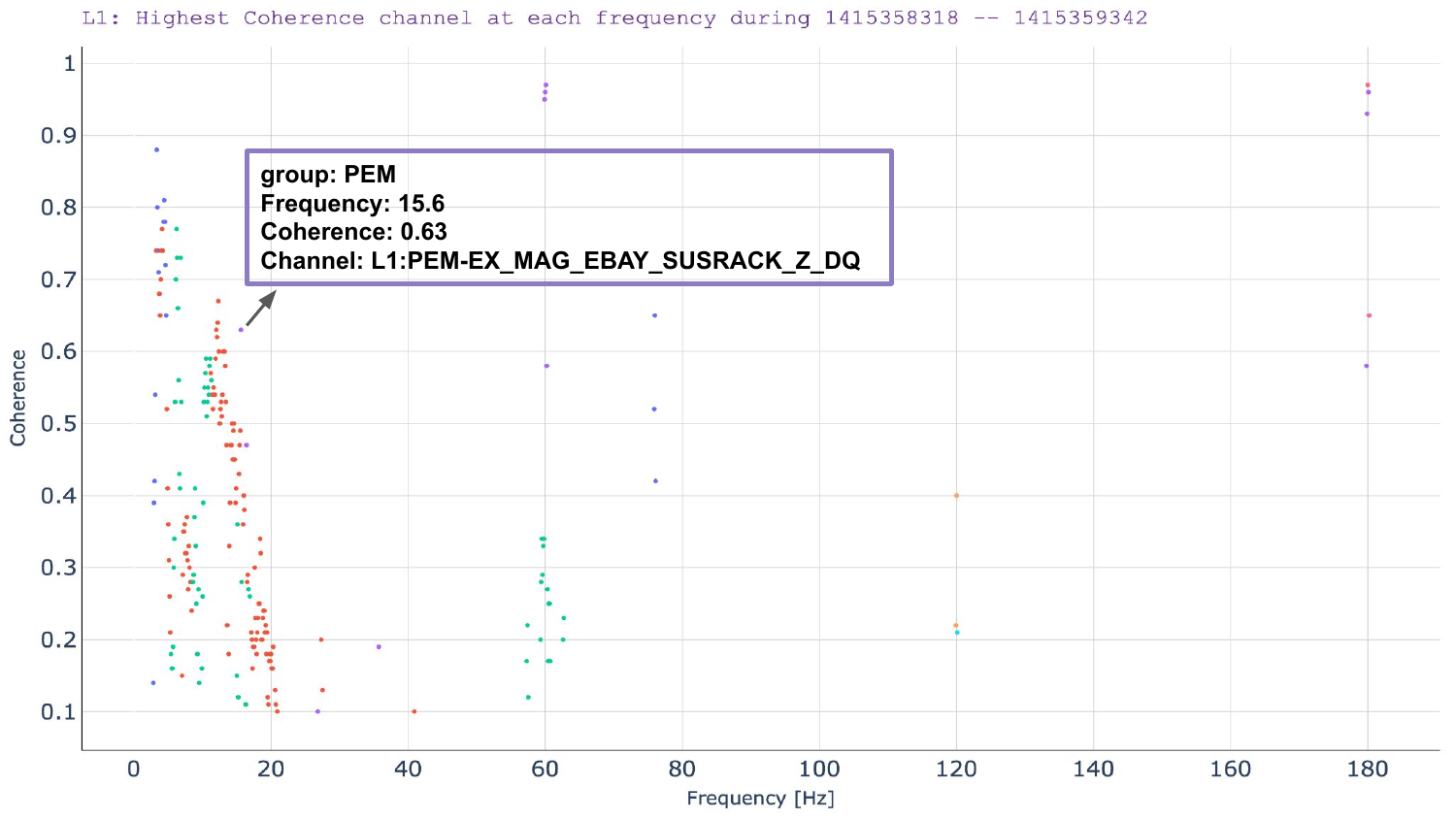}
    \caption{This plot shows the maximum coherence at each frequency between 0 and 200 Hz with a resolution of 0.1 Hz between LLO strain channel and auxiliary channels between GPS time 1415358318 and 1415359342. A coherence threshold of 0.1 was applied. These plots are generated for each run of the {\tt CohMon} analysis. In its interactive web version, hovering over the data points in the plot displays the channel name and other relevant information as shown here.}
    \label{fig:LLO_Coherence_Nov9}
\end{figure*}

Ground based gravitational-wave detectors are highly dynamic and complex systems. They are subject to changing environmental and instrumental conditions that can create new or modify existing noise couplings. This necessitates continuous monitoring of the coherence between the auxiliary and strain data streams. Narrow spectral artifacts also known as lines adversely impact the gravitational-wave search sensitivity and the LIGO Detector Characterization group employs multiple tools such as {\tt BruCo}, {\tt Fscan}, {\tt STAMP-PEM} and {\tt StochMon} to study and monitor these lines \cite{PhysRevD.101.042003, Fscan, Meyers:2018nyo, Stochmon}.
In this section we discuss a new tool called {\tt CohMon} that we developed for measuring the coherence between the strain data and hundreds of ``safe" auxiliary channels once every hour of the observing time for both detectors \cite{cohmon-presentation}. The primary motivation behind Python based {\tt CohMon} is witness channel identification so that this information can be fed to noise cleaning tool {\tt DeepClean} for immediate noise subtraction. 
The channels to be analyzed and the cadence of analysis are all configurable parameters. For each (auxiliary) channel analyzed, the coherence value measured up to the Nyquist frequency (half the sampling frequency) for the given channel 
is saved in an output file.
As part of monitoring the coherence, {\tt CohMon} generates diagnostic plots that show the channel with the highest coherence at each frequency analyzed in the band $0$--$200~\mathrm{Hz}$. Figure \ref{fig:LLO_Coherence_Nov9} shows a representative of such plot. The data products of {\tt CohMon} can be used for downstream analyses and for any noise cleaning tools, as we discuss in section~\ref{sec:Denoising}. 

\subsection{Noise stationarity}

Changes in the instrument and surrounding environmental conditions can lead to changes in the couplings between the auxiliary channels and the gravitational-wave strain.
Understanding the time-variation of couplings may assist in detector commissioning work. It is also an essential measurement when performing software-based denoising as the relevance of auxiliary channels may change vastly over time. 
In its current version, {\tt CohMon} records the coherence between the auxiliary channels and $h(t)$ on an hourly basis. This choice of time-stride reflects a balance between the computational overhead and the time scale over which we have empirically established noise behavior to be subject to significant changes. In this approach, any variability of noise can be identified only if it occurs over periods longer than the coherence baseline of an hour that we use.
Variability in noise couplings as identified by {\tt CohMon} may also correlate with transient-noise artifacts in $h(t)$, often referred to as ``glitches" if these glitches occur repeatedly over several hours, which is the case with scattered light glitches.
One of the most frequent sources of transient noise in the LIGO Livingston detector is scattered light \cite{Soni:2023kqq, Glanzer:2022avx}. 
During O4, slow scattering of light due to increased ground motion in the $0.1$--$0.5~\mathrm{Hz}$ band is adversely impacting the $h(t)$ data quality in the $10$--$50~\mathrm{Hz}$ band \cite{LIGO:2024kkz}. Radio frequency reflected light ports in the length sensing and control subsystem (LSC REFL RF) are good witnesses of this slow scattering transients. During the November 7-11, 2024 data-taking, we notice an increase in microseismic activity near the LIGO Livingston detector, which leads to an increase in slow scattering glitches in the $h(t)$. This change in rate of noise is reflected in the change in coupling of these auxiliary channels as captured by {\tt CohMon} and shown in the left plot of figure \ref{fig:LLO_Coherence_Nov9}. The right plot shows the rate of slow scattering glitches in $h(t)$ as identified by Gravity Spy on Nov7, Nov 9 and Nov 11~\cite{Zevin:2016qwy, Soni:2021cjy}.

\begin{figure}
  \centering
  \begin{minipage}{0.99\linewidth}
    \includegraphics[width= \textwidth]{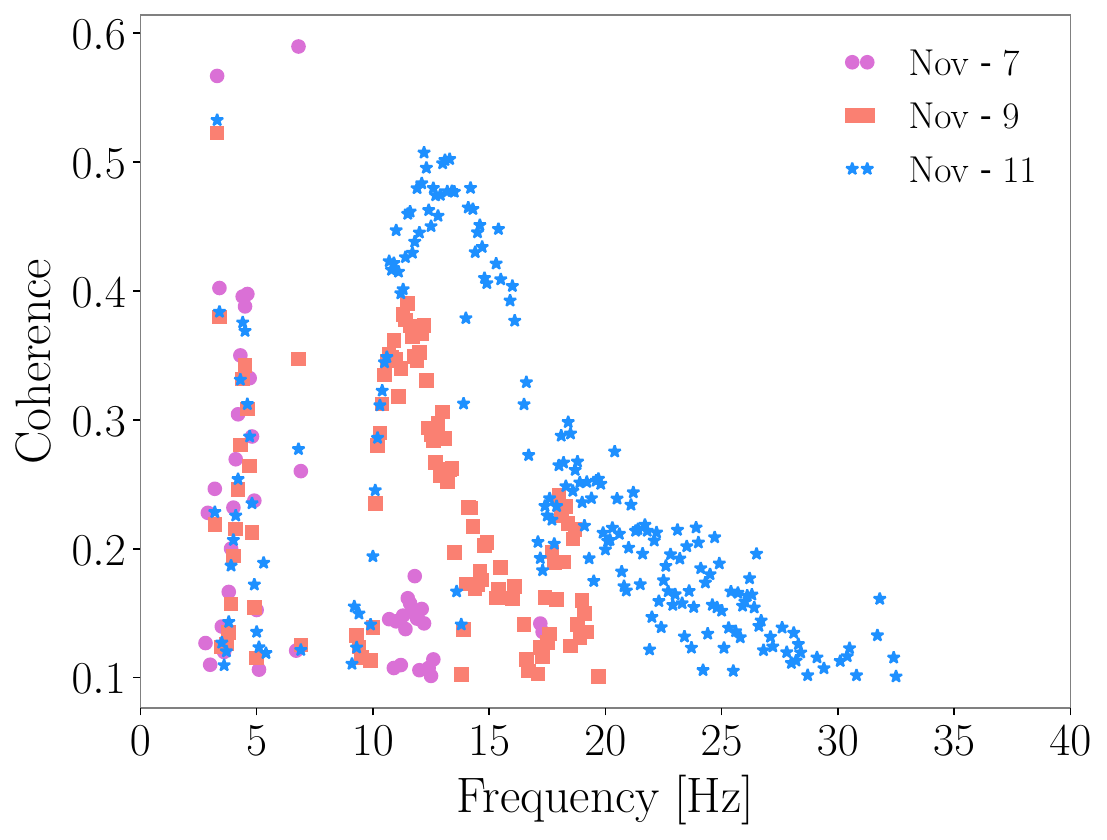}
    \label{fig:coh_llo_nov}
  \end{minipage}
  \hfill
  \begin{minipage}{0.99\linewidth}
         \includegraphics[width =\textwidth]{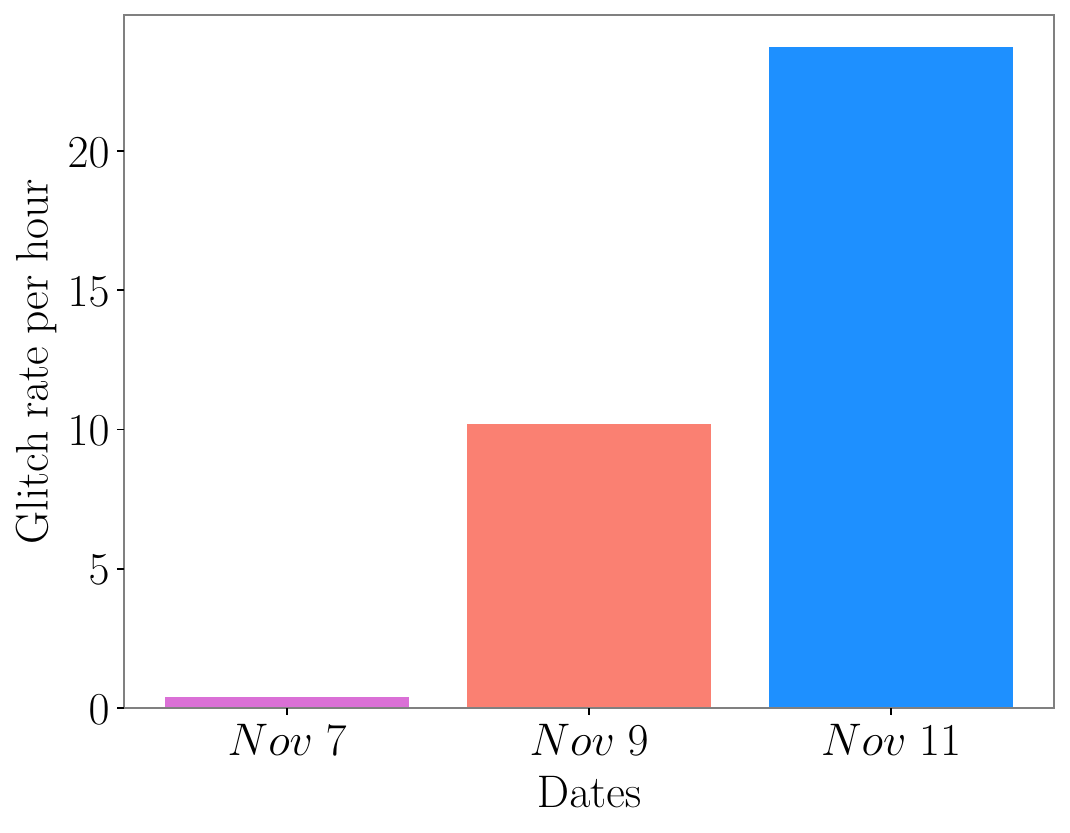}
         \label{fig:gspy_llo}
  \end{minipage}
\caption{\emph{Top}: Coherence between the length sensing and control radio frequency reflected light port (LSC REFL RF) photodiode and $h(t)$ at LIGO Livingston is seen increasing during November 7-11 2024 measurements. This auxiliary channel is a good witness of stray light noise in the strain data. \emph{Bottom}: Over the same time span, an increased rate of transient noise is classified as ``slow scattered light" by GravitySpy \cite{Zevin:2016qwy,robinet2020omicron}.}
\label{fig:LLO_glitch_coh}
\end{figure}

\begin{figure}
  \centering
  \begin{minipage}{0.99\linewidth}
         \includegraphics[width= \textwidth]{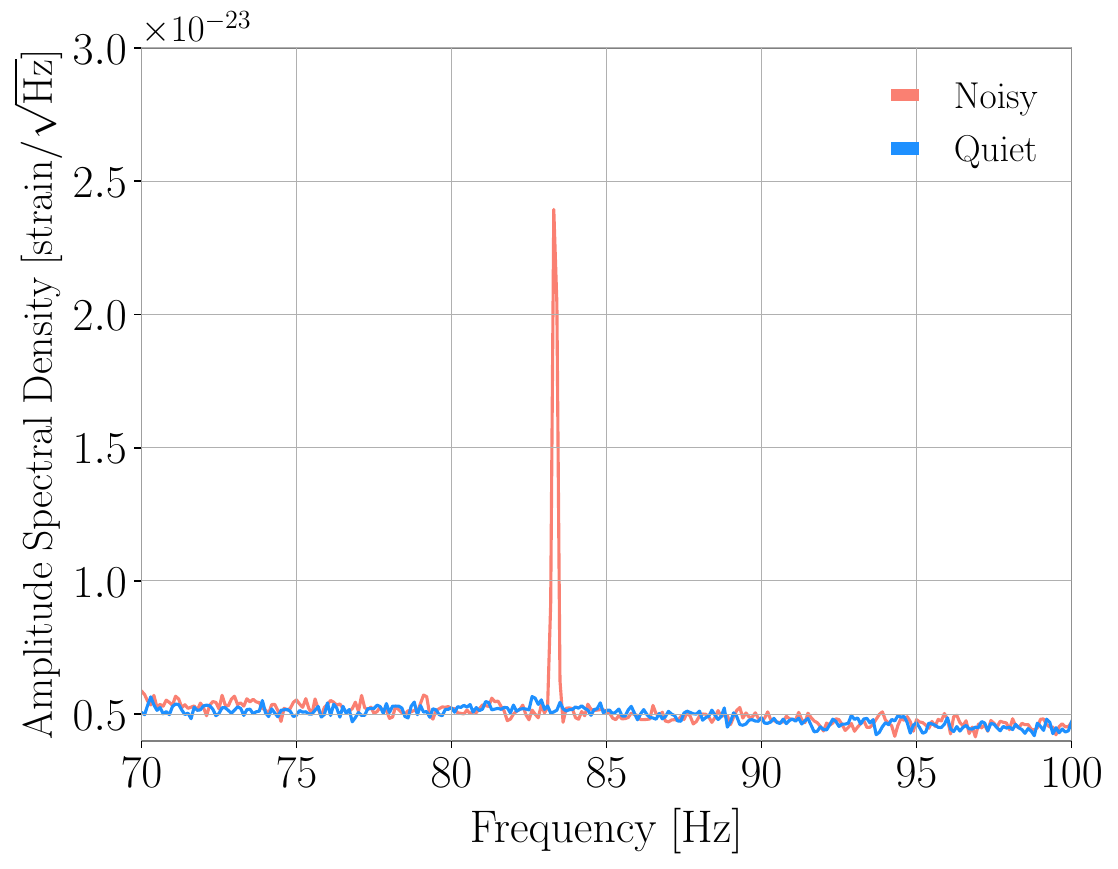}
         \label{fig:83Hznoise}
  \end{minipage}
  \hfill
  \begin{minipage}{0.99\linewidth}
         \includegraphics[width =\textwidth]{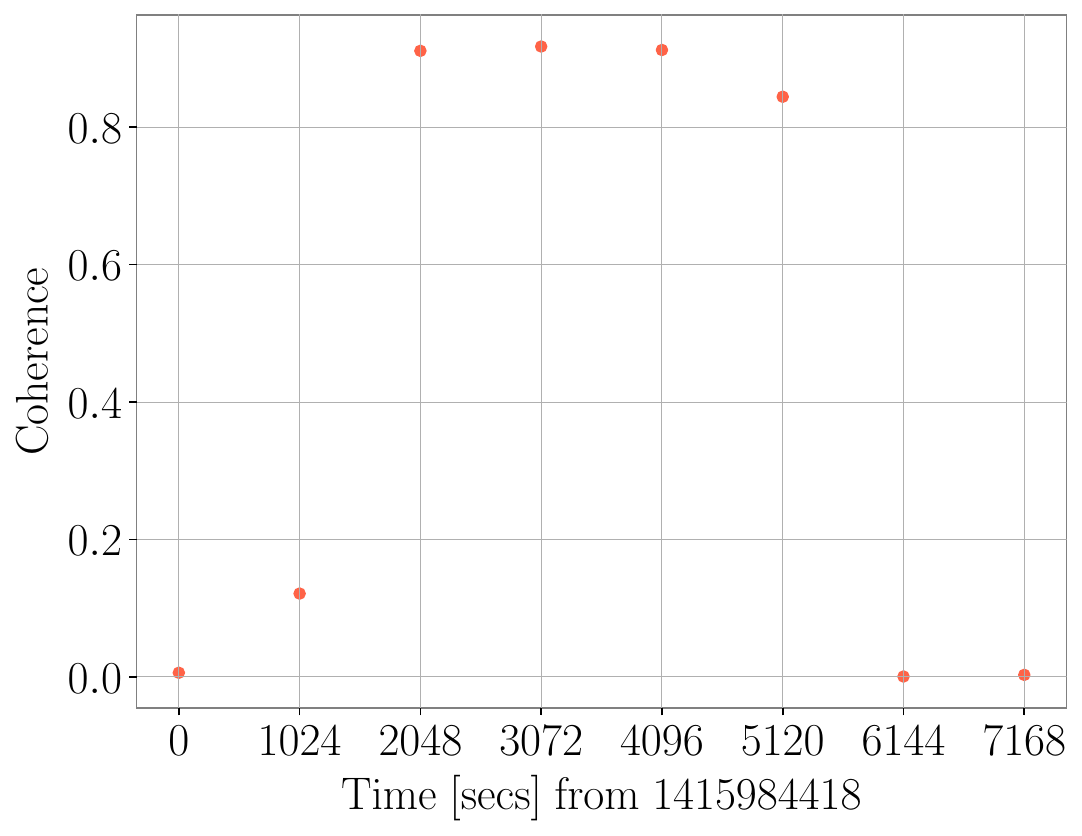}
         \label{fig:accnoise}
  \end{minipage}
  \caption{\emph{Top}: The peak at $83.3$ Hz in the strain amplitude spectral density appears after commissioning work at the LIGO Livingston mid-station. \emph{Bottom}: Coherence at 83.3 Hz between mid-station accelerometer and $h(t)$ measured for every 1024 seconds from GPS time 1415984418 or 5 pm UTC on Nov 18 2024.}
    \label{fig:83Hzacc}
\end{figure}

\label{sec:83.3Hz}
Statistically significant variation in coherence as a function of time may also be the unintended result of 
detector commissioning work or other environmental changes on the detectors.
Coherence monitoring can be used to connect the appearance of noise both over time as well as with any detector commissioning work on auxiliary components of the detector. This can be helpful in our understanding of how changes in one part of the detector can impact other.
For example, detector commissioning work at the X-arm mid-station started driving the baffle resonance at LIGO Livingston around 5:30 pm UTC on November 18, 2024 ~\cite{74088}. Baffles are used to reduce the stray light noise coupling in the LIGO detectors \cite{baffles}.  
This baffle resonance close to $83.3$ Hz increased noise in $h(t)$ around the same frequency as shown in the top panel of figure~\ref{fig:83Hzacc}. The increased motion was detected by the mid-station accelerometer. This increase in coherence close to 83.3 Hz between the accelerometer and $h(t)$ is picked up by {\tt CohMon}: the bottom panel in figure~\ref{fig:83Hzacc} shows the coherence measured by {\tt{CohMon}} between the accelerometer and the strain data. The first two data points show the low coherence before the baflle was driven at its resonant frequency, the next four data points are during the increased baffle motion~\cite{74088}. Once this problem was resolved, the coherence goes down again as shown by the last two data points in this figure. 

Understanding and reducing the noise couplings in the detector is one of the major studies gravitational-wave instrument scientists undertake; continuous coherence monitoring with {\tt CohMon} aids these efforts. At the same time the identification and ranking based on coherence values of possible noise witness channels using {\tt CohMon} provides the basis for noise subtraction in software.
This is a powerful approach whenever a hardware fix is not attainable, or the data are already acquired. 

\section{Noise Removal}
\label{sec:Denoising}
Since long-duration environmental and instrumental noise is known to also exhibit non-linear couplings, using an artificial neural network approach for denoising is expected and has been demonstrated to improve sensitivity with respect to traditional noise-cleaning techniques. This non-linear coupling and denoising was the focus of the work in reference~\cite{PhysRevResearch.2.033066, Saleem:2023hcm}. It shows how non-linear couplings during LIGO's O3 run that appeared as sidebands around the 60 Hz mains line can be successfully subtracted. Furthermore, previous work in reference~\cite{PhysRevResearch.2.033066, Saleem:2023hcm} demonstrated that the \texttt{DeepClean} algorithm does not affect astrophysical signals.

We make explicit use of the \texttt{DeepClean} algorithm~\cite{Saleem:2023hcm, PhysRevResearch.2.033066}, which consists of a convolutional neural network that follows an encoder-decoder structure. The information from the witness channels is encoded to a lower-dimensional latent space from which the decoder regresses the combined noise budget. Trainable parameters are updated by minimizing the power spectral density (PSD) ratio between the original strain and the strain minus the predicted noise budget. We refer the reader to reference~\cite{PhysRevResearch.2.033066} for more details regarding the architecture and training procedure. 

\begin{figure*}
\centering
\includegraphics[width=.98\linewidth]{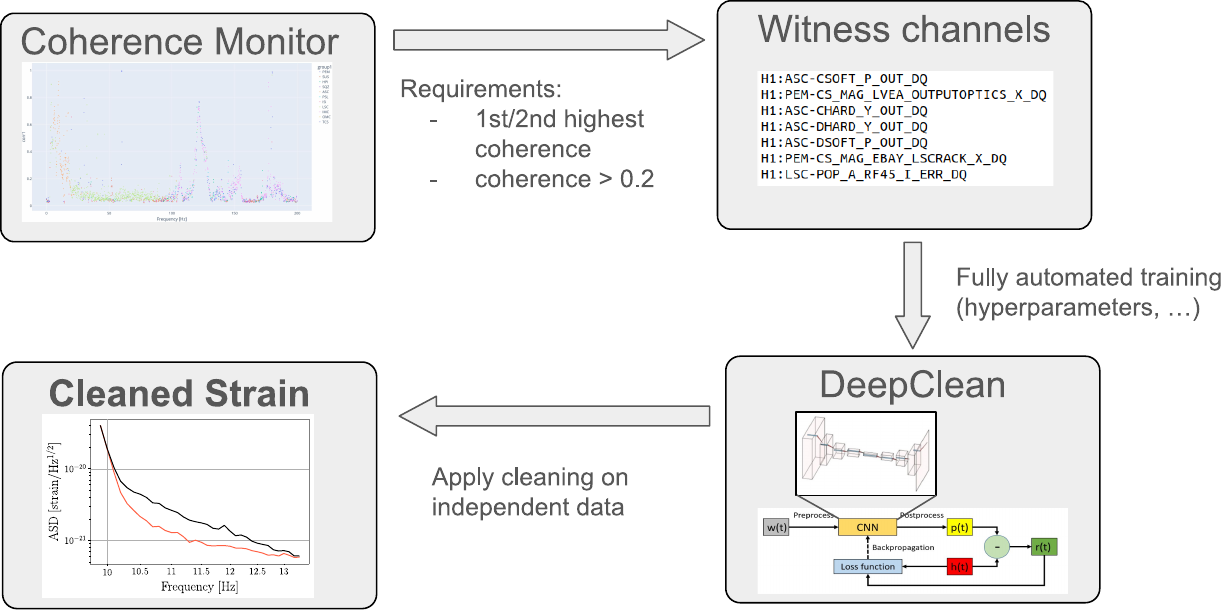}
\caption{Schematic overview of the {\tt Coherence DeepClean} ({\tt CDC}) pipeline. The {\tt DeepClean} panel is adopted from ~\cite{Saleem:2023hcm, PhysRevResearch.2.033066}.}
\label{fig:Pipeline}
\end{figure*}

Before this work, {\tt DeepClean} has been demonstrated to clean only broadband noise due to beam jitter noise in O2  as well as narrow line noise related to the 60\,Hz power line since relevant noise couplings needed to be externally identified~\cite{PhysRevResearch.2.033066,Saleem:2023hcm}. The lack of an algorithmic basis for identifying suitable witness channels prevented the application of {\tt DeepClean} for noise removal beyond beam jitter and 60\,Hz power line noise. With the coherence monitor described in Section~\ref{sec:Coherence_Monitor}, we close this gap. Our end-to-end, combined pipeline aims to identify and remove a wide variety of noises. We refer to this pipeline as {\tt Coherence DeepClean} ({\tt CDC}). It is shown schematically in figure~\ref{fig:Pipeline} and is made up of the following building blocks.
\newline
\noindent
{\bf{Coherence Monitoring:}} we monitor the coherence between the strain channel and all auxiliary channels considered ``safe", in total approximately 900 channels per detector, using 1024\,s of data and a Fast Fourier Transform window of 10\,s.
\newline
\noindent
{\bf{Witness Channel Selection:}}
from the coherence monitor output and for a given frequency interval, we extract the list of witness channels that are highest correlated to the strain. Frequencies are scanned in steps of 0.1\,Hz meaning that for an interval of 3\,Hz the list contains 31 witness channels. Similarly, we extract a list for the second-highest correlated witness channel per frequency in the considered frequency band. We combine the lists, remove duplicate channels, and require at least a peak coherence of 0.2 between each witness channel and the strain in the given frequency band. This choice is made to optimize the number of witness channels (thus the computation complexity of the training and the model size) versus the cleaning performance. Depending on the noise removal problem, we find the number of selected witness channels to vary between zero and 35, with zero indicating that no suitable witness channel monitoring the noise is found and subsequent denoising is not possible. The upper limit of 35 is what we find experimentally by creating witness channel lists on the data we analyze.
\newline
\noindent
{\bf{Training:}}
we train the {\tt DeepClean} algorithm on the same 1024\,s long data segment used for coherence monitoring, using only the selected witness channels as inputs. We split the data $90\%-10\%$ between training and validation, respectively, with the first consecutive 922\,s of data being used for training and the rest for validation. Testing is performed on 4096\,s of data directly following the training and validation interval. To avoid potential biases, the training data cannot be cleaned with the model trained in the respective time interval. To avoid data gaps for subsequent analysis, for \texttt{CDC}, we use a model trained on earlier data to clean the training interval.
\newline We apply an eighth-order Butterworth bandpass to choose frequencies in the targeted band. Additionally, all data is normalized before applying the machine learning algorithm. We divide data into overlapping segments where each segment is 8\,s long with an offset of 0.25\,s between data segments. While we fix the batch size to 32, the optimal learning rate depends strongly on the number of witness channels and their respective couplings. 
Therefore, we update the original {\tt DeepClean} algorithm, adding a learning rate tuning that reduces the learning rate by a factor of 0.2 in case the PSD ratio between the original and cleaned strain does not decrease over two training epochs. We find the right choice of the learning rate to be essential for learning various noise features of the data. Choosing a learning rate that is too small, the algorithm fails to learn broadband noise features and only focuses on subordinate spectral line noise contributions. We choose the initial learning rate to be $4\times10^{-3}$, which is subsequentially decreased to learn both broadband and spectral line noise.
\newline
\noindent
{\bf{Inference:}}
we test the performance of our trained machine learning model on a 4096\,s long time series directly following the training and validation interval. Similarly to the training data, we bandpass and normalize the inputs. Predictions are derived for the same data segment length used in training (8\,s), while the overlap between segments is 4\,s. The predictions from overlapping segments are combined by applying Hanning windows to the data.

The hyperparameters of our neural network architecture are derived following hyperparameter tuning. Various analysis parameters (e.g., time-stride, frequency band, coherence threshold, number of channels) have been optimized using a grid search. They are all configurable to match computational, latency, and resolution constraints, although our findings have no strong dependence on the choices.

Depending on the exact cleaning problem, the data-taking period, and the number of selected witness channels, selecting suitable witness channels, training, and inference as described above takes between several to up to 15 minutes utilizing a single NVIDIA TITAN X Pascal GPU on the International Gravitational-Wave Observatory Network (IGWN) computing grid.

\section{Results}
\label{sec:Results}
\begin{figure}
\centering
\includegraphics[width=.99\linewidth]{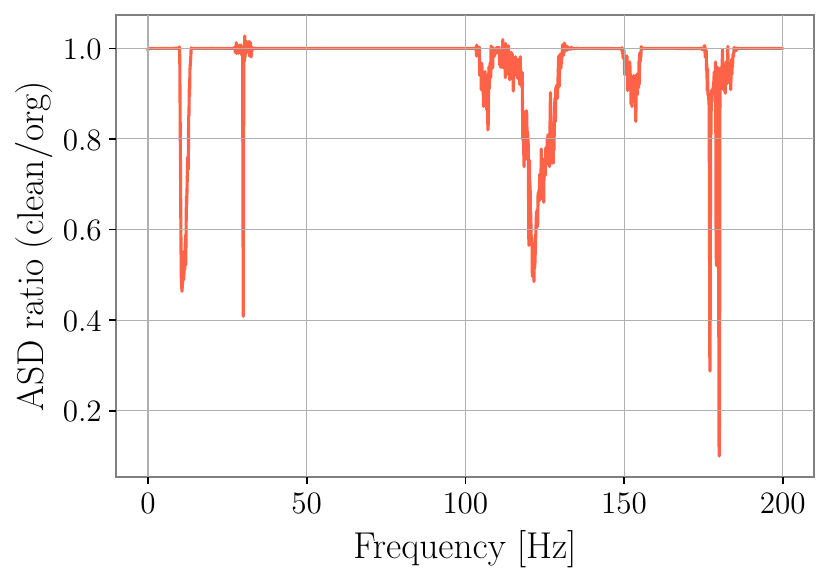}
\caption{Ratio of strain amplitude spectral density (ASD) before and after denoising with the {\tt CDC} pipeline for data taken during the fourth observing run (O4) on September 10, 2023 (GPS time 1378403243) by the Hanford detector. We apply the pipeline to six distinct frequency bands where suitable witness channels have been identified with the coherence monitor {\tt CohMon}. Ratios slightly above and below one can occur due to statistical fluctuations.}
\label{fig:Ratio}
\end{figure}

\begin{figure*}
  \centering
  \begin{minipage}{0.49\linewidth}
  \centering
  \includegraphics[width=.99\textwidth]{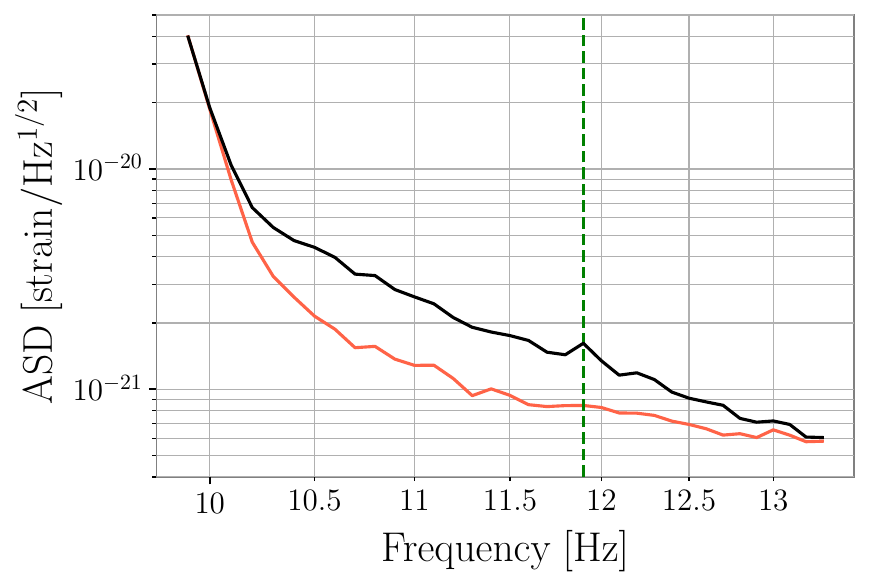}
  \end{minipage}
  \hfill
  \begin{minipage}{0.49\linewidth}
  \centering
  \includegraphics[width=.99\textwidth]{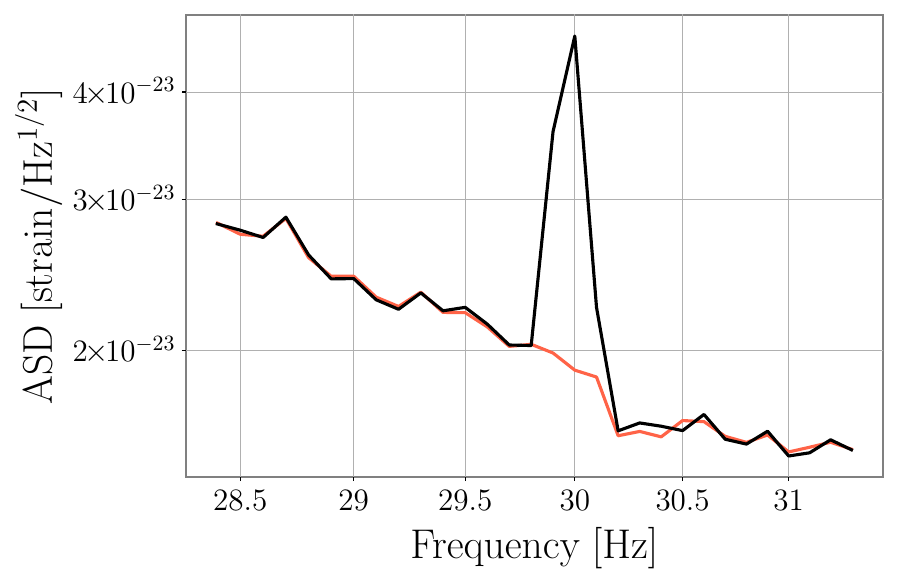}
  \end{minipage}
  \begin{minipage}{0.49\linewidth}
  \centering
  \includegraphics[width=.99\textwidth]{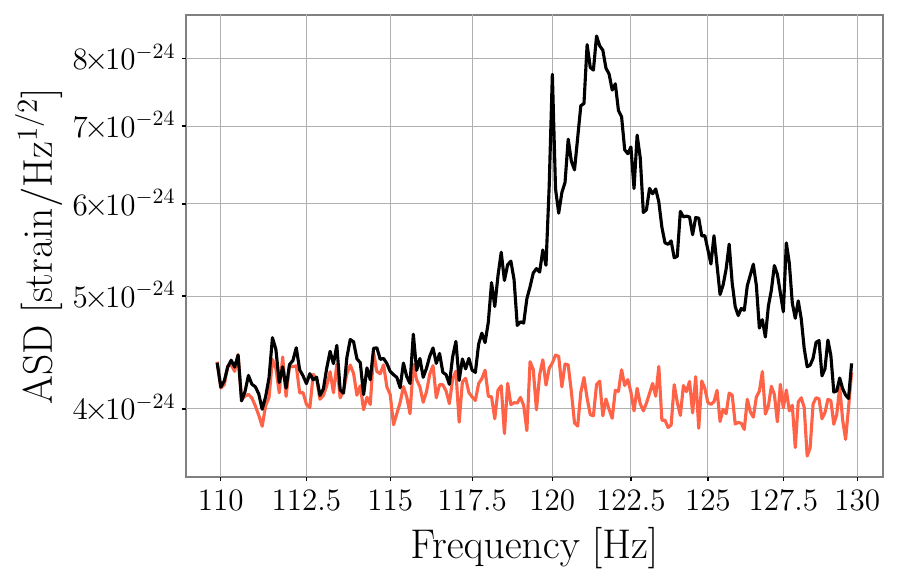}
  \end{minipage}
  \hfill
  \begin{minipage}{0.49\linewidth}
  \centering
  \includegraphics[width=.99\textwidth]{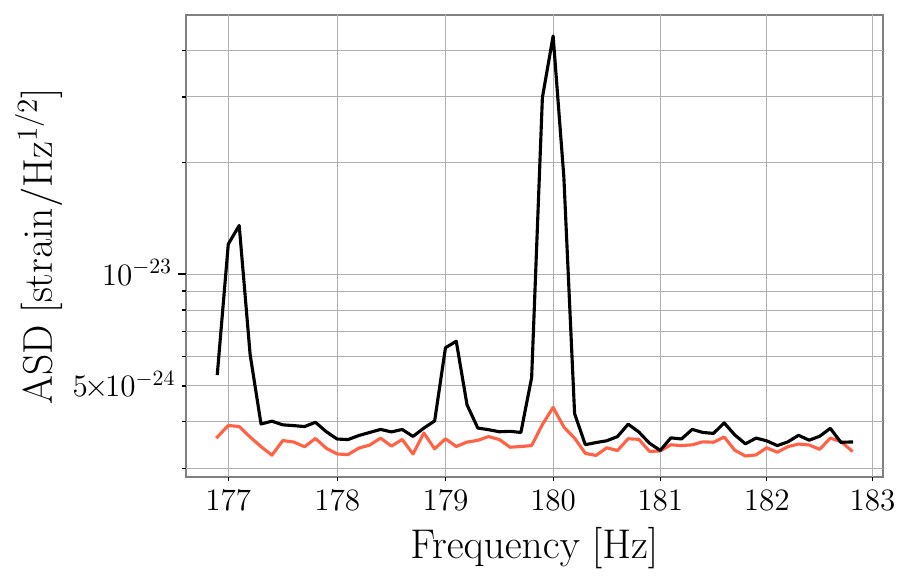}
  \end{minipage}
\caption{Amplitude spectral density (ASD) before (black) and after (red) denoising for dedicated frequency bands using data taken with the LIGO Hanford detector. The plots show the variety of noise that can be removed with the pipeline. \emph{Top left}: broadband noise, \emph{Top right}: spectral lines, \emph{bottom}: combination. The green dashed line marks 11.9\,Hz to highlight the characteristic noise feature successfully linked to the output of two magnetometers.}
\label{fig:NoiseCharacteristic}
\end{figure*}

We obtain all results reported here by applying the \texttt{CDC} pipeline to the calibrated gravitational-wave strain signal channel 
\verb|GDS_CALIB_STRAIN|.
Tests show that findings do not change significantly when using instead the calibrated strain with low-latency line subtraction 
\verb|GDS_CALIB_NOLINES|.
Using data taken by the LIGO Hanford detector starting September 10, 2023 (GPS time 1378403243), we identify six frequency bands with suitable amounts of witnessed noise: 10-13\,Hz, 28.5-31.5\,Hz, 104-109\,Hz, 110-130\,Hz, 150-155\,Hz and 177-183\,Hz. The amplitude spectral density (ASD) ratio of the original to the cleaned signal strain can be found in figure~\ref{fig:Ratio}. In figure~\ref{fig:NoiseCharacteristic}, we show the ASD for particular frequency bands, demonstrating that by applying the {\tt CDC} pipeline, we can subtract noise with varying characteristics from broadband noise to the removal of single spectral lines or the combination of both. 
The characteristics of the selected witness channels vary depending on the frequency range. For the LIGO Hanford detector, below 20\,Hz we find dominant contributions from channels monitoring alignment sensing and control of the optics and magnetometers monitoring the local magnetic field. At higher frequencies, mostly between 100 and 200\,Hz, we also find channels related to the input mode cleaning to show high coherence. Exemplary, we show the strong interplay between witness channel identification and denoising performance in figure~\ref{fig:NoiseCharacteristic} (top left).
We find the output of two magnetometers located in the laser and vacuum equipment area and the electronics bay area to have high coherence at 11.9\,Hz. If these channels are included as inputs and the training is conducted as described, we find the characteristic noise at 11.9\,Hz to be removed. If these channels are excluded from the algorithm inputs, the noise persists.

\begin{figure}
\centering
\includegraphics[width=1.1\linewidth]{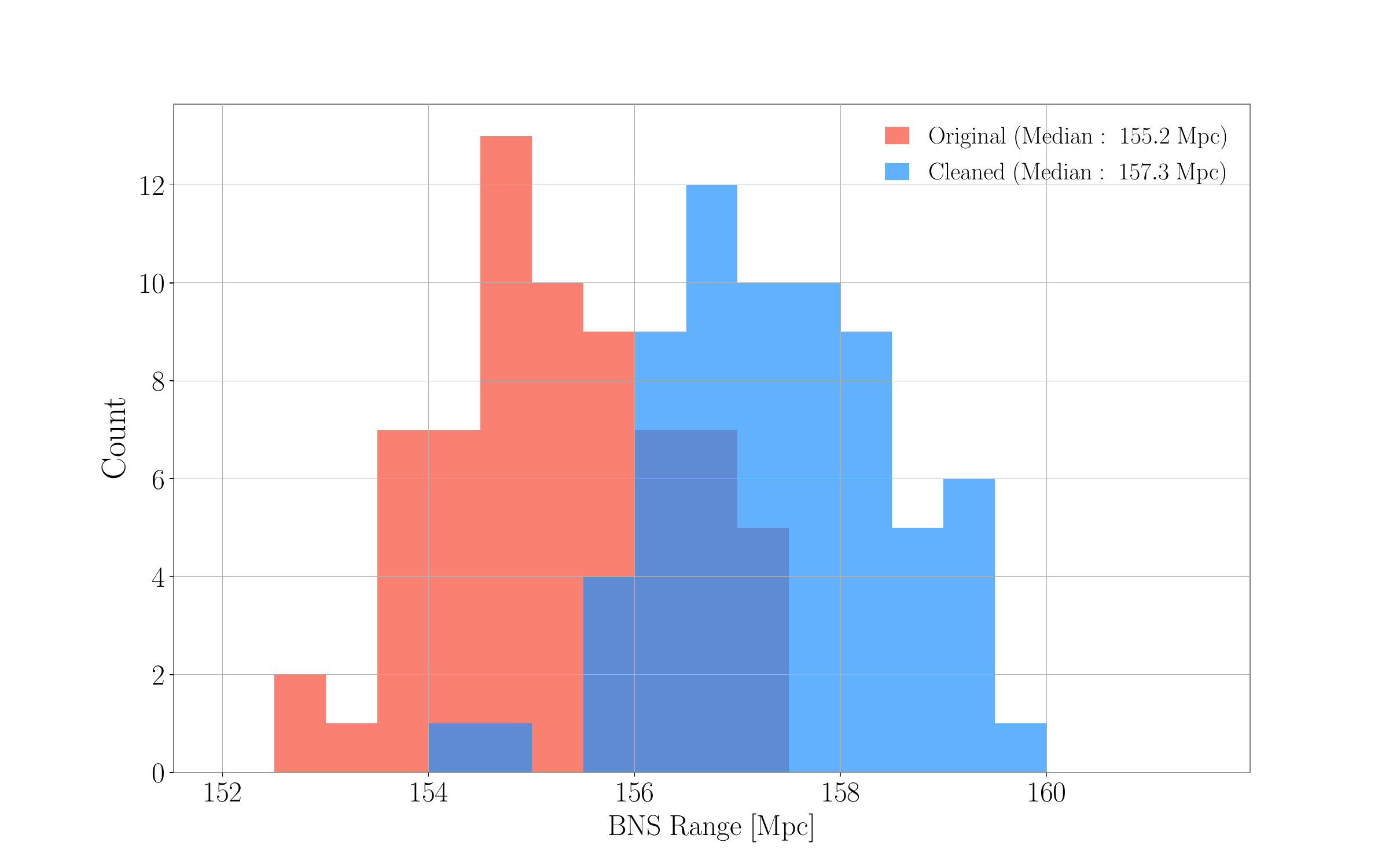}
\caption{Binary Neutron Star (BNS) Range calculated on 60\,s long data segments taken with the LIGO Hanford detector on September 10, 2023 (GPS time 1378403243), before and after applying the {\tt CDC} pipeline. In total, the (original) cleaned time series is 4096\,s long, resulting in 68 entries per time series in the histogram.}
\label{fig:BNS}
\end{figure}

The reduction of the detectors' ASD through this denoising process directly impacts the instruments' astrophysical reach.
We compute the inspiral range, i.e., the sky- and polarization-averaged distance within which two binary neutron stars with 1.4 solar masses each will be detected with a signal-to-noise ratio of eight or above \cite{Chen:2020zoq}. The results for the individual 60\,s long time segments can be found in figure ~\ref{fig:BNS}. We find that the mean inspiral range after the cleaning for the LIGO Hanford detector increases by $1.4 \%$, which translates to an increase in the total sensitive astrophysical volume of $4.3\%$.

\begin{figure}
  \centering
  \begin{minipage}{0.99\linewidth}
  \centering
  \includegraphics[width=.99\textwidth]{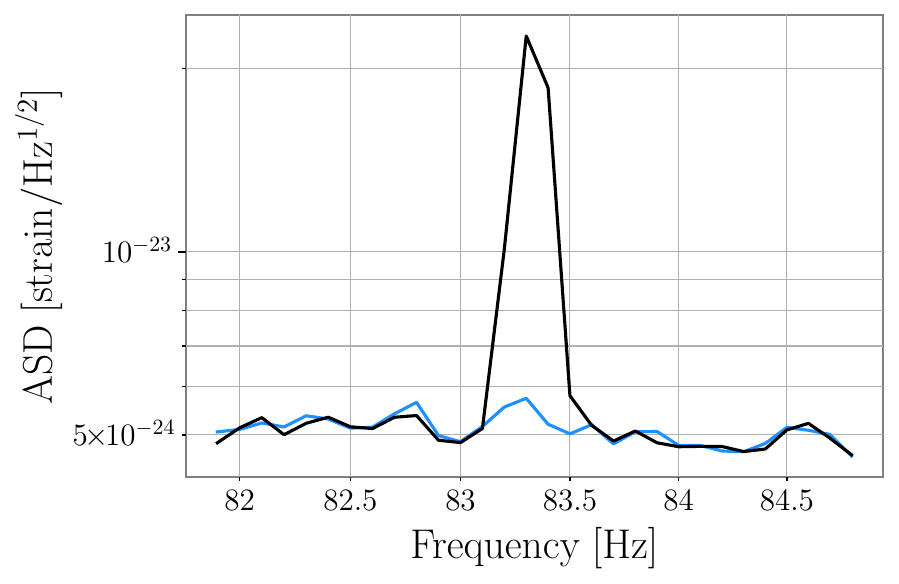}
  \end{minipage}
  \hfill
  \begin{minipage}{0.99\linewidth}
  \centering
  \includegraphics[width=.99\textwidth]{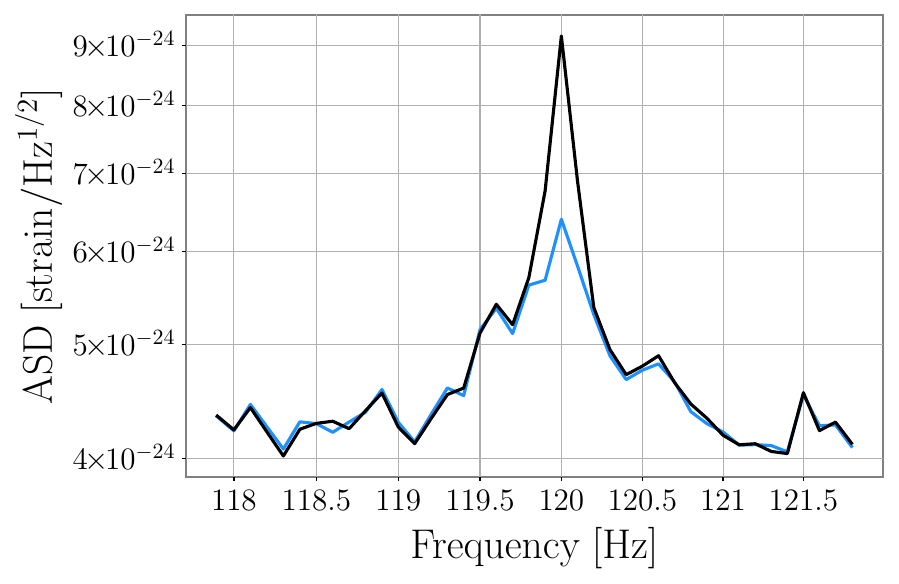}
  \end{minipage}
\caption{Amplitude spectral density (ASD) before (black) and after denoising (blue) for dedicated frequency bands. Data was taken with the LIGO Livingston detector on November 18, 2024 (GPS time 1415997186) (left) and July 20, 2023 (GPS time 1373846418) (right). Both lines correspond to excess noise that was only observed over several hours.}
\label{fig:ASD_Livingston}
\end{figure}

While, in general, we observe significantly less monitored low-frequency noise in data taken with the LIGO Livingston detector, the {\tt CDC} pipeline is particularly suitable to target changing noise conditions. The performance of the {\tt CDC} pipeline in removing this shorter-duration stationary noise using data taken by the LIGO Livingston interferometer on November 18, 2024 (GPS time 1415997186) and July 20, 2023 (GPS time 1373846418) is shown in figure~\ref{fig:ASD_Livingston} (top) and figure~\ref{fig:ASD_Livingston} (bottom), respectively. The baffle resonance that leads to the excess noise at 83.3\,Hz and how we use the coherence monitor to identify it are discussed in detail in section~\ref{sec:83.3Hz}. Demonstrating the ability of removing noise that only persists over several hours or days, we show the full potential of the {\tt CDC} pipeline as a universal tool for automated monitoring of experimental conditions, identifying new noise sources, and removing noise directly in software. 

\section{Summary and outlook}
\label{sec:Summary}
We developed an automated coherence monitor to analyze ground-based gravitational-wave detector environmental and instrumental noise couplings. Given the outputs obtained with the coherence monitor, we select suitable witness channels for denoising the gravitational-wave strain, utilizing a neural network ({\tt DeepClean}) to predict the combined noise budget. The pipeline, which we refer to as {\tt Coherence DeepClean}, or {\tt CDC}, allows us to subtract disparate noise contributions from spectral lines, broadband and composite noise for the first time. Given the interferometer and the start time of the train and test segment, the {\tt CDC} pipeline automatically identifies frequency bands and auxiliary channels and produces cleaned data frames. While we demonstrate using this {\tt CDC} pipeline for denoising data taken with the Advanced LIGO detectors, it is universal. There are ongoing efforts for {\tt DeepClean} applications in denoising Advanced Virgo and KAGRA data and results obtained will be reported in separate publications.
While the {\tt CDC} pipeline improves our understanding of the interferometers and may also help to mimic noise couplings in hardware, it is a first important step torwards fully autonomous denoising of gravitational-wave detector data in software, adapting to changing noise conditions with minimal human intervention and minimal turn-around times.  

\begin{acknowledgments}
\label{sec:ack}
CR acknowledges support under a Postdoc.Mobility fellowship from the Swiss National Science Foundation (SNF) (Project No.\ P500PT\_222340). SS and EK acknowledge support from the National Science Foundation (NSF) under award PHY-1764464 and PHY-2309200 to the LIGO Laboratory.
MS acknowledges the support from Weinberg Institute for Theoretical Physics at the University of Texas at Austin.
MS and MC acknowledge NSF support under grant number PHY-2308862.
All authors acknowledge NSF support to the Institute on \textit{Accelerated AI Algorithms for Data-Drive Discovery} (A3D3) under award PHY-2117997.

This research was undertaken with the support of the LIGO computational clusters.
The material presented is based upon work supported and data collected by the LIGO Laboratory which is a major facility funded by the National Science Foundation.
\end{acknowledgments}

\appendix


\bibliographystyle{unsrtnat}
\bibliography{apssamp}
\end{document}